\documentclass[9pt,twoside]{article}
\usepackage[a4paper,top=3cm,bottom=3.5cm,left=2.5cm,right=2.5cm]{geometry}
\usepackage{cite}
\usepackage{subfigure}
\usepackage{amssymb}
\usepackage{lineno}
\usepackage[english]{babel}
\usepackage{palatino,url}
\usepackage[utf8x]{inputenc}
\usepackage{amsmath}
\usepackage{amssymb}
\usepackage{listings}
\usepackage{graphicx}
\usepackage[rightcaption]{sidecap}
\usepackage{textcomp}
\usepackage{array} 
\usepackage{booktabs} 
\usepackage{multirow} 
\usepackage{caption} 
\usepackage{floatflt}
\usepackage{subfigure}
\usepackage{rotating} 
\usepackage{scrextend}
\usepackage{multirow}
\deffootnote[1em]{1em}{1em}{\textsuperscript{\thefootnotemark}\,}
\usepackage[colorinlistoftodos]{todonotes}

\title{A Multi-agent approach for \textit{in silico} simulations of micro-biological systems}
\author{Proverbio D., Gallo L., Passalacqua B., Pellegrino J., Maggiora M.}
\date{}

\begin{document}

\maketitle

\begin{abstract} 
\noindent Using a Multi-agent systems paradigm, the present project develops, validates and exploits a computational \textit{testbed} that simulates micro-biological complex systems, namely the aggregation patterns of the social amoeba \textit{Dyctiostelium discoideum}. We propose a new design and implementation for managing discrete simulations with autonomous agents on a microscopic scale, thus focusing on their social behavior and mutual interactions. Then, the dependence on the main physical variables is tested, namely density and number of amoebas; in addition, we analyze the robustness of the dynamics against various noise sources. Along with these results, we suggest a methodology for further studies that make use of our validated model.
\end{abstract}

\section{Introduction}
\label{Sec1}
The behavior of the social amoeba \textit{D. discoideum} \cite{kessin2,romeralo} represents an archetype of swarm intelligence. It has inspired models of distributed control \cite{bonabeau} and a bio-inspired model for problems of decentralized gathering \cite{strmecki2005developmental}, cell signaling and chemotaxis \cite{bonner,devreotes,williams2010dictyostelium}. In fact, if food sources (bacteria or agar) are available, each cell lives and proliferates individually; once starving, though, a multitude (up to 100,000 cells) of single amoebas starts sending pulses of 3,'5'-Cyclic Adenosine Monophosphate (cAMP) molecules \cite{morris} to its surroundings \cite{chen,robertson,van,day} which, after being received and transduced by membrane receptors \cite{othmer1998oscillatory}, let \textit{D. discoideum} migrate via chemotaxis \cite{durston,futrelle,bonner2}. The final stage of the gathering process is the formation of a multicellular slug organism \cite{darmon,fey}.\\

 \begin{figure}[h!]
  \centering  
  \includegraphics[scale=0.15]{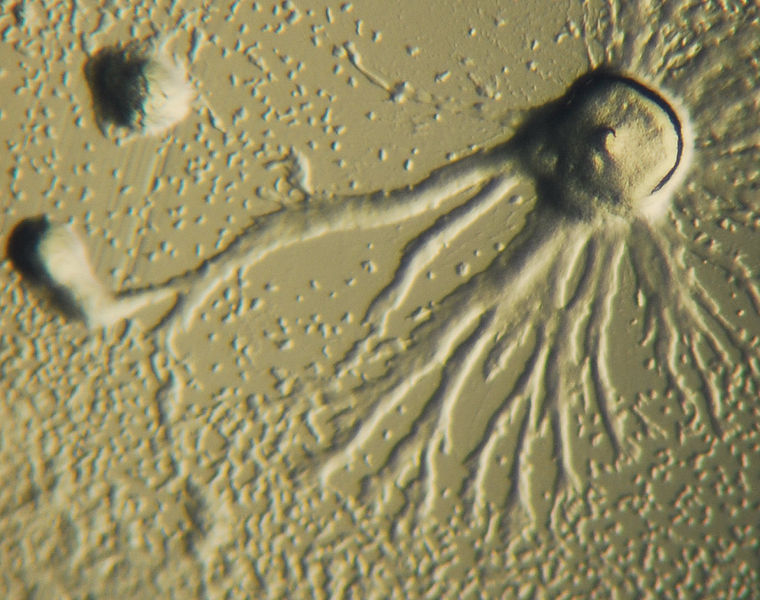}
  \caption{A gathering of \textit{D. discoideum} \cite{photo}. Cells can be seen migrating (some individually, and some in streams) toward a central body \cite{kessin}.}
  \label{0a}
   \end{figure}

\noindent To inquire \textit{D. discoideum} dynamics, the most common modeling approaches tackle such systems via continuous diffusion equation \cite{gierer,dallon,bakerpdes}, eventually solved with Monte Carlo methods \cite{palsson,oss1996spatial} or, more recently, with in-lattice discrete simulations \cite{nagano,pineda}. Other than that, the last years IT developments made it possible to analyze the microscopic processes underlying self organization of a colony while relying on a hybrid continuous-discrete modeling approach (still on a macroscopic scale) \cite{fates,nagano2}.\\

\noindent Besides them, the main contribution of the present project is to follow a \textit{Multi-Agent Systems} computational approach to a) build an agent architecture that is consistent with biological knowledge; b) define metrics to quantitatively study the gathering process; c) validate the model by means of comparison with well-established results and by generation of predictive data.

\section{Methodology}
The change of modeling paradigm was justified as the aforementioned methods from literature mainly describe the dynamics from a macroscopic perspective \cite{grune}. However, they are not capable of inquiring the emergence of complex patterns from microscopic behaviors \cite{bakerpdes}; at the same time, it is impossible to gauge the impact of individual stochasticity or failure on the overall evolution of the system, thus resulting in poor resolution on the system control side \cite{iglesias2010control}. Therefore, rather than following said approaches, we exploited a bottom-up paradigm, namely that of \textit{Multi-Agent Systems} (MAS). \\
An \textit{agent} is an individual ``computer system that is capable of independent action on behalf of its user or owner [...] A \textit{multi-agent system} is one that consists of a number of agents, which interact with one another, typically by exchanging messages through some computer network infrastructure" \cite{woolridge}.
As a special feature, these models allow to account for intracellular decision processes which are triggered by biochemical cell-cell or cell-matrix interactions \cite{galle}. In addition, they provide natural candidates for modeling the evolution and pattern formation of large multi-cellular systems \cite{parhizkar} since they tie cellular properties to macroscopic behavior on the population level \cite{drasdo,kennedy,zeghida}. \\
Other than that, the agent-based paradigm allows for a natural management of communication issues among individuals: specifically, we were able to model the cAMP pulses without using an instantaneous \textit{point to point} messaging \cite{singhal} nor diffusing waves \cite{cantrell2004spatial} (as it was done in most of continuous reaction-diffusion models \cite{fates,nagano,dallon}). On the contrary, modeling the signals as discrete traveling packages reduces the computational complexity. Given \textit{N} the number of agents, we obtain $\mathcal{O}$(\textit{N}) instead of state-of-the-art $\mathcal{O}$ (\textit{N $\log^2$N}) for fractional diffusion equations \cite{wang2010}, thus allowing better scaling for large populations. Validity of such vectorial approach was checked by comparing generated data (qualitative and quantitative ones) with those in literature that make use of diffusing waves or geometrical approaches \cite{fates,dallon,palsson,elliott,vidal}. Finally, since the dynamics is generated during each simulation run, such models may be used as testbeds to control hypothesis and to generate predictive data \cite{saetzler}.\\

\noindent As a result, we obtain a finite set of (logical) behavioral rules that lead to an emergence of collective synchronization, and we verify if and under which conditions the global self-aggregation is obtained. In particular, thanks to the chosen scale, both the microscopic and macroscopic variables that define the model, namely the number and the density of cells, were studied and their role assessed. To end with, predictions about the system behavior are made, discussed and used to assess its robustness.

\section{Design and Implementation}
\begin{figure}[h!]
  \centering
  \includegraphics[scale=0.25]{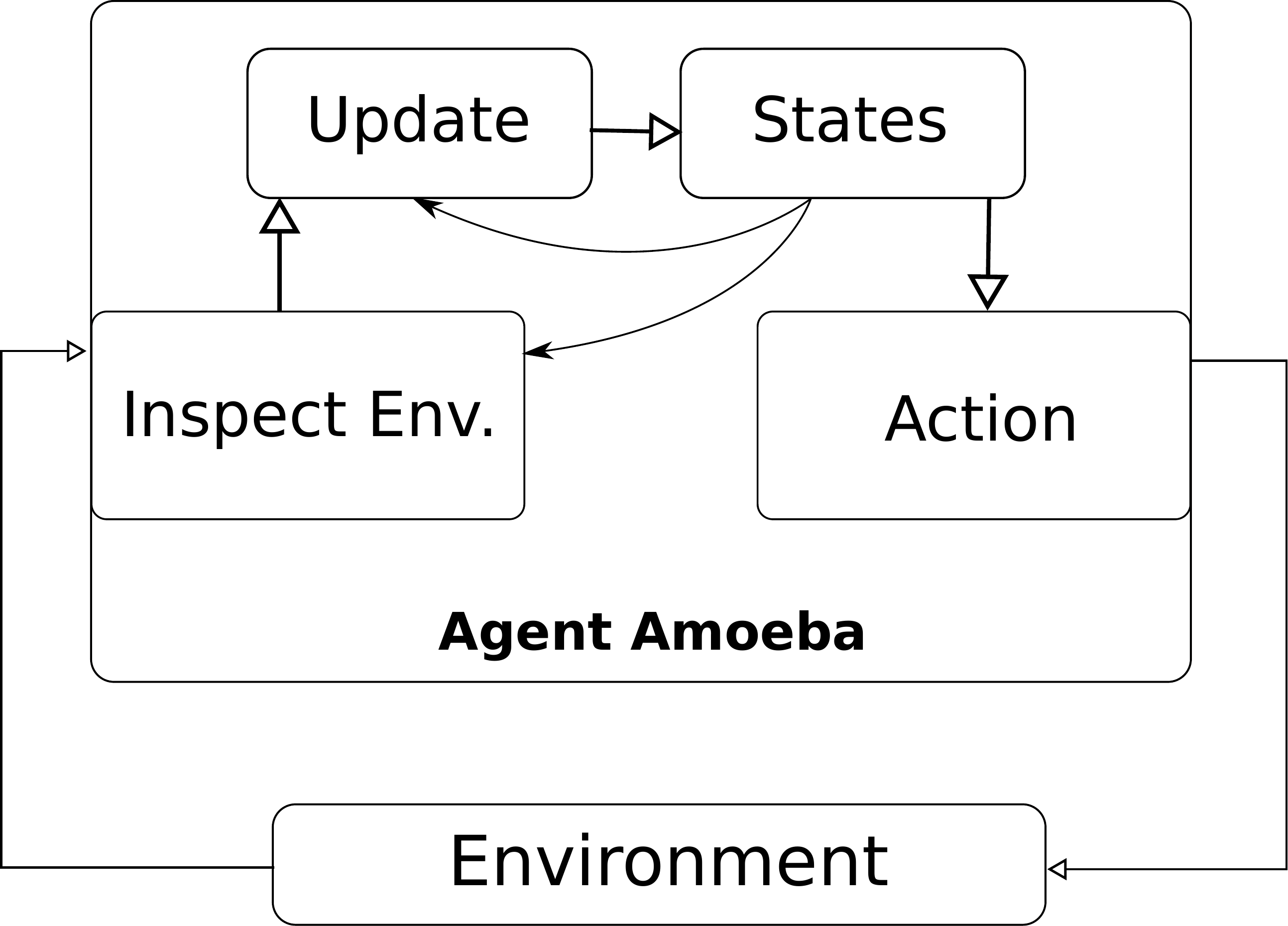}
   \caption{\small Schematic representation of the Amoeba agent architecture and its environment. Given the set of configurations \{E$_{g}\}$ of the environment $\mathrm{Env}$, the agent is a tuple $\mathrm{Am} = \langle inspect, internal, action \rangle$ where ``Internal" is a set of features \{update, states\} that are not directly interfacing with the environment (those that distinguish the agent from a purely reactive one). In particular, States = \{S$_{j}$\} is the set that defines the internal peculiarities of each agent and that selects the other methods; Inspect = \{I$_{i}^{j}$: E$_{g} \mapsto \mathrm{per}$\} sees E$_{g}$ and generates an internal perception $\mathrm{per}$; Update = \{U$_{l}^{j}$: $\mathrm{per} \mapsto S_{j}$\} may update the states according to the perceptions; Action = \{A$_{k}^{j}$: S$_{j} \mapsto E_{g'}$\} is the set of feasible actions on $\mathrm{Env}$ according to the state the agent is in.}
 \label{0b}
 \end{figure}

Our model consists of three main agent species: a rigid, non-toroidal, dynamically evolving 2-D \cite{bonner1998} grid (the environment), amoeba agents (simulating single cells) and vectorial cAMP packages (carrying signal). \\
After a thorough literature search (see Sec. \ref{Sec1} as well as \cite{alcantara,nan,willard,jowhar}), Amoeba agents are implemented with individual behaviors embedded in a complex architecture, as they are not purely reactive, yet keep information about the environment and their past history (see Fig. \ref{0b}). Note that all Amoeba agents are alike (no pace-maker cells were introduced at a first approximation \cite{gross}). They are given two specific states: wandering $\mathbb{W}$ (there is food, so the amoeba is eating) and starving $\mathbb{S}$ (the amoeba is \textit{not} finding food, so it begins the aggregation). As we are only interested in looking for gathering patterns, we didn't add ``aggregated'' or ``slug'' states \cite{coates2001cell}. Amoebas are capable of the following: while their inner state is $\mathbb{W}$, they wander and eat the bacteria; as soon as their state becomes $\mathbb{S}$, they isotropically shoot signals (with a $\pm 20^\circ$ random smearing $\delta_1$) and, if they are reached by a traveling cAMP agent, they orientate according to the direction of the just-absorbed signal and move on (a smearing $\delta_2 = 10 \%$ on the trajectory is also considered) \cite{jowhar}. Since the main focus is on the aggregation patterns, we neglect the complex inner synchronization mechanisms within each cell \cite{kim2009robustness}; on the contrary, we rely on the demonstrated robustness of said mechanisms \cite{kim2007stochastic}. Inertial effects are also neglected, as cells and micro-organisms typically exist in low-Reynold's number environments \cite{purcell}.\\

\noindent After setting the dimension $\mathrm{dim}$, the environment is represented by a square domain $\mathrm{Env} = \langle x_{k};y_{k} \rangle$ (where $k = 0 \dots \mathrm{dim} \in \mathbb{N}$), each cell of it being assigned a scalar value (whose evolution might be ruled by appropriate parameters) that represents the food sources, the presence of which makes the \textit{in-silico} experiment more realistic and better adherent to wet-lab biological observations, where several factors may affect amoebas behavior. Formally, at a given time $t \in \mathbb{N}$, a food source ($m$-th bacteria) is $b[t]: (\{x_{m},y_{m}\} \in \mathrm{Env}, t) \mapsto b \in \mathbb{N}$.\\
\noindent cAMP agents are sent vectorial signals, representing discrete packages of chemicals: they are shot radially by amoebas and they travel across the environment, with a speed $\mathbf{v_{cAMP}}$ whose module is equal to that of the field mean diffusion speed, and are eventually absorbed by other amoebas or the borders. Consequently, what varies is not precisely the diffusive gradient, but the probability associated to finding a cAMP agent with defined direction: from Agent A perspective, absorbing an Agent B's signal is equal to $P^{A}(\{ I^{starving}_{B-cAMP}: E \mapsto \mathrm{per} \}) \propto \frac{1}{r^2}$ where $r = v_{cAMP} \cdot \hat{t}$ is the length of the cAMP trajectory ($\hat{t}$ corresponds to the elapsed time between emission from A and absorption by B). \\

\noindent To implement the code, we chose the MAS platform GAMA \cite{gama}, that lets us a) handle the agents consistently with the desired design; b) manage the parameters easily; c) have a user-friendly interface showing the whole environment that allowed us to inspect what was happening in real time and to obtain informative animations.

\section{Launching an experiment}
An experiment is defined by the choice of the agents parameters, the run of repeated simulations and the measurements of various quantitative variables.

\subsection{Parameters and variables}
\label{params}
The discussed model contains different free parameters, whose value is to be set. These were chosen to be biologically consistent \cite{alcantara,dallon,fey,nan,willard}, although scaled (note that a simulation cycle equals 1 min). Therefore the values of reference are: diameter of amoeba $D$ = 1.8 units (while the real typical dimension is at order of 18$\mu$m); speed of amoeba $v_{A}= 0.3$ $\frac{\mathrm{units}}{\mathrm{cycle}}$ \cite{mccann}, which corresponds to the average speed leading to the ``visual'' starting of an aggregate after six hours \cite{fey}; diffusion speed of cAMP molecules $v_{c}= 0.9$ $\frac{\mathrm{units}}{\mathrm{cycle}}$; cAMP shooting time $t_{s}$ = 10 cycle, this latter indicating the mean period between two cAMP shootings.
The reference value for $\mathrm{dim}$ (edge of the environment) is 100 units = 55.5 $\times D$ = 1 mm. Note that, from now on, the measure unit \textit{units} will be implied.\\
The model manages ``thermal noise'' parameters as well: $P_{G}$ represents the growth probability $P(b[t+1] \leftarrow b[t]+1)$ of bacteria in the environment, namely the probability that a bacteria cell increases its scalar value by one at the given time; $P_{A}$ is the ``agitation probability'' $P_{A} = P(\{I_{i}^{j}:E_{g} \nrightarrow \mathrm{per}\})$, that represents a biological source of detection failure, namely an asymmetric binding receptor occupancy that, during chemotaxis, leads to a noisy input signal or signal detection failure \cite{van2007}. When interpreting the whole system from a statistical mechanics perspective, the latest parameters represent the contribution on the thermal $\beta$, therefore the term ``noise''. During the process of validating the model these values were set arbitrarily; their impact on the dynamics is studied shortly afterwards.\\

\noindent As stated previously, we are interested in understanding the dependence of the model on the variables density ($\rho$) and number ($N$) of amoebas (macroscopic and microscopic variables, respectively), which are mutually related by 
\begin{equation}
\rho = \frac{N \cdot A}{\mathrm{dim}^{2}}
\label{rho}
\end{equation}
with $A$ being the area of a single amoeba. The reference density ranges from $\frac{80 \cdot A}{100^{2}}$ up to $\frac{600 \cdot A}{100^{2}}$ and greater, the same magnitude as the (typically biological) range of $(\frac{10000 \cdot A}{cm^{2}} \div \frac{100000 \cdot A}{cm^{2}})$ as reported in \cite{dallon,fey}. Therefore, in order to tune $\rho$, we need to act over $\mathrm{dim}$ alone, holding $N$ fixed, in order to separate the contribution of the microscopic variable from the macroscopic one. After that, we adequately rescale the ``spatial'' parameters, namely v$_{A}$, v$_{c}$, $P_{G}$. On the contrary, as $D$ is already involved in $A$, we do not scale it. From now on, ``$\cdot A$'' will be implied.

\subsection{Simulations}
\begin{figure}[h]
\centering
\textbf{Example: evolution of a single simulation}\\[0.2cm]

\begin{minipage}[c]{0.3\textwidth}
\includegraphics[scale=0.28]{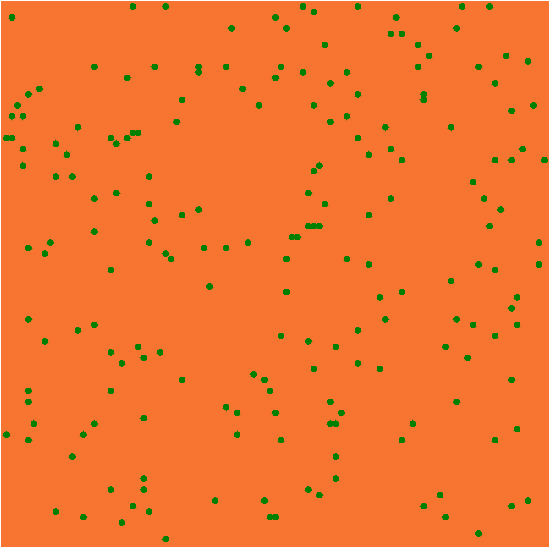}
\renewcommand{\figurename}{Fig.}
\caption{Initial setting of an average simulation.}
\label{1a}
\end{minipage}
\hspace{0.4cm}
\begin{minipage}[c]{0.3\textwidth}
\includegraphics[scale=0.28]{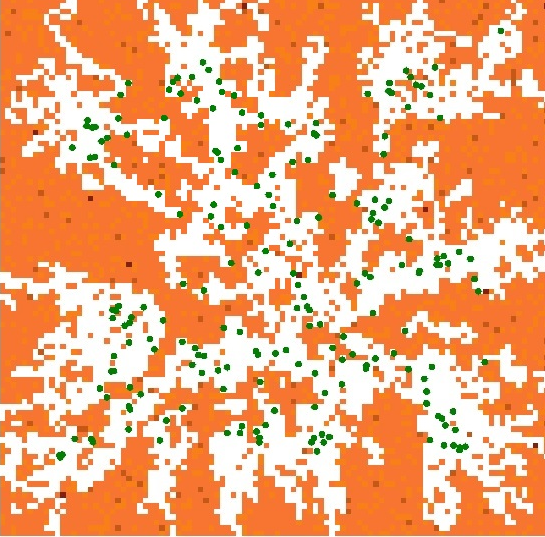}
\renewcommand{\figurename}{Fig.}
\caption{Evolution and movement.}
\label{1b}
\end{minipage} 
\hspace{0.4cm}
\begin{minipage}[c]{0.3\textwidth}
\includegraphics[scale=0.28]{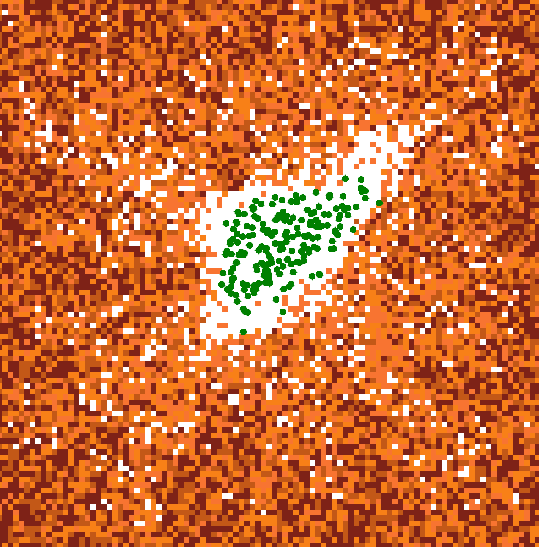}
\renewcommand{\figurename}{Fig.}
\caption{Aggregation and simulation end.}
\label{1c}
\end{minipage}
\caption*{\small Three steps of a simulation with a relatively small number of amoebas (for clarity reasons). Setting: green dots are scattered amoeba cells, the orange landscape depicts a uniform bacteria distribution (Fig. \ref{1a}). Run: amoeba agents move dynamically across the environment, while bacteria growth (brown pixels) is ruled by $P_{G}$. White areas are without bacteria, as amoebas ate them in previous cycles. At each cycle, the new position of an agent is calculated according to its direction and its speed. cAMP agents are not shown, although traveling across the environment at the same time as amoebas (Fig. \ref{1b}). Aggregation: amoeba cells form an aggregate. As we didn't implement an ``aggregated'' state, the slug is hold tight solely by mutual movement along signals and not by a potential field representing cellular adhesion (Fig. \ref{1c}).}
\end{figure}

Given an initial random cell configuration and a certain bacteria distribution in the domain, a run is characterized by the steps depicted in Fig. \ref{1a}, \ref{1b}, \ref{1c}.\\

\noindent When it comes to qualitative features of a certain system such as pattern formation, the first an experimentalist does is to look for visual similarities \cite{mackay1978computer}. As a consequence, the behaviors observed in animations have been confirmed by means of comparison with biological snapshots, videos (Fig. \ref{0a}, \cite{dicriscio}) and results from previous models \cite{nagano,nagano2,dallon,palsson,elliott,fates,dallon2011understanding}.
Finally, repeated simulations for each set of parameters and variables are run in order to obtain statistics.

\subsection{Measurements}
\begin{figure}[h!]
  \centering
  \includegraphics[scale=0.25]{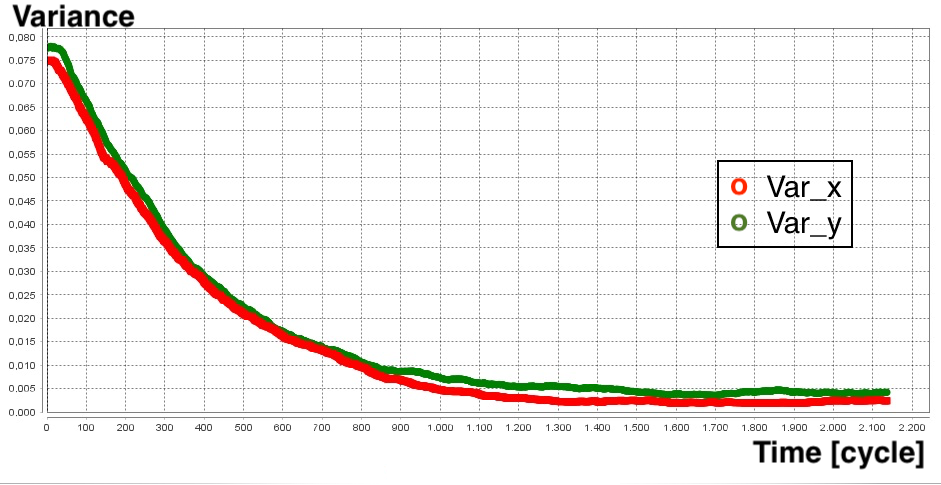}
   \caption{Evolution of the marginal dispersions $\mathrm{var}_{X}^{i}$ and $\mathrm{var}_{Y}^{i}$ during time, as shown by GAMA graphic interface.}
 \label{2}
 \end{figure}

Being the first step to validate the present model, we focused on the aggregation rate. The order parameter is the time $t$ defined by simulation cycles. The chosen quantitative variable to measure the rate is the marginal posterior variance along the two axis x and y. Being $\{x_{n}, y_{n}\}$ the coordinates of the $n$-th amoeba in $\mathrm{Env}$, for each $i$-th simulation we measure
\begin{equation}
\mathrm{var}^{i}(t) = \frac{1}{\mathrm{dim}^2}\frac{1}{N}\sum_{n=1}^N ||E(x_n,y_n)(t)-(x_n,y_n)(t)||
\end{equation}
were $\bar{x}(t)$ is the mean value at each cycle $t$. This way we get information about the global behavior towards aggregation around a specific center as the \textit{dispersion} of the colony. Note that, thanks to the rotational invariance of $\mathrm{Env}$, $\frac{1}{\sqrt{2}}\mathrm{var}^{i}(t) = \mathrm{var_X}^{i}(t)$, so that we can directly use the latter for further analysis (it is lighter to be directly calculated and saved by GAMA software). In Fig. \ref{2} is shown the evolution of the dispersions $\mathrm{var}_{X}^{i}$ and $\mathrm{var}_{Y}^{i}$ during time, for one simulation, as depicted by GAMA interface. \\
Having repeated measures, it is natural to evaluate the mean value and the associated statistical error as the standard deviation of $\mathrm{var}_{X}^{i}(t)$: ($\mathrm{var}_{X}(t) \pm \sigma_{var_{X}}(t)$). This method will be repeated for each experiment.

\section{Analysis}
\begin{figure}[h!]
  \centering
  \includegraphics[scale=0.35]{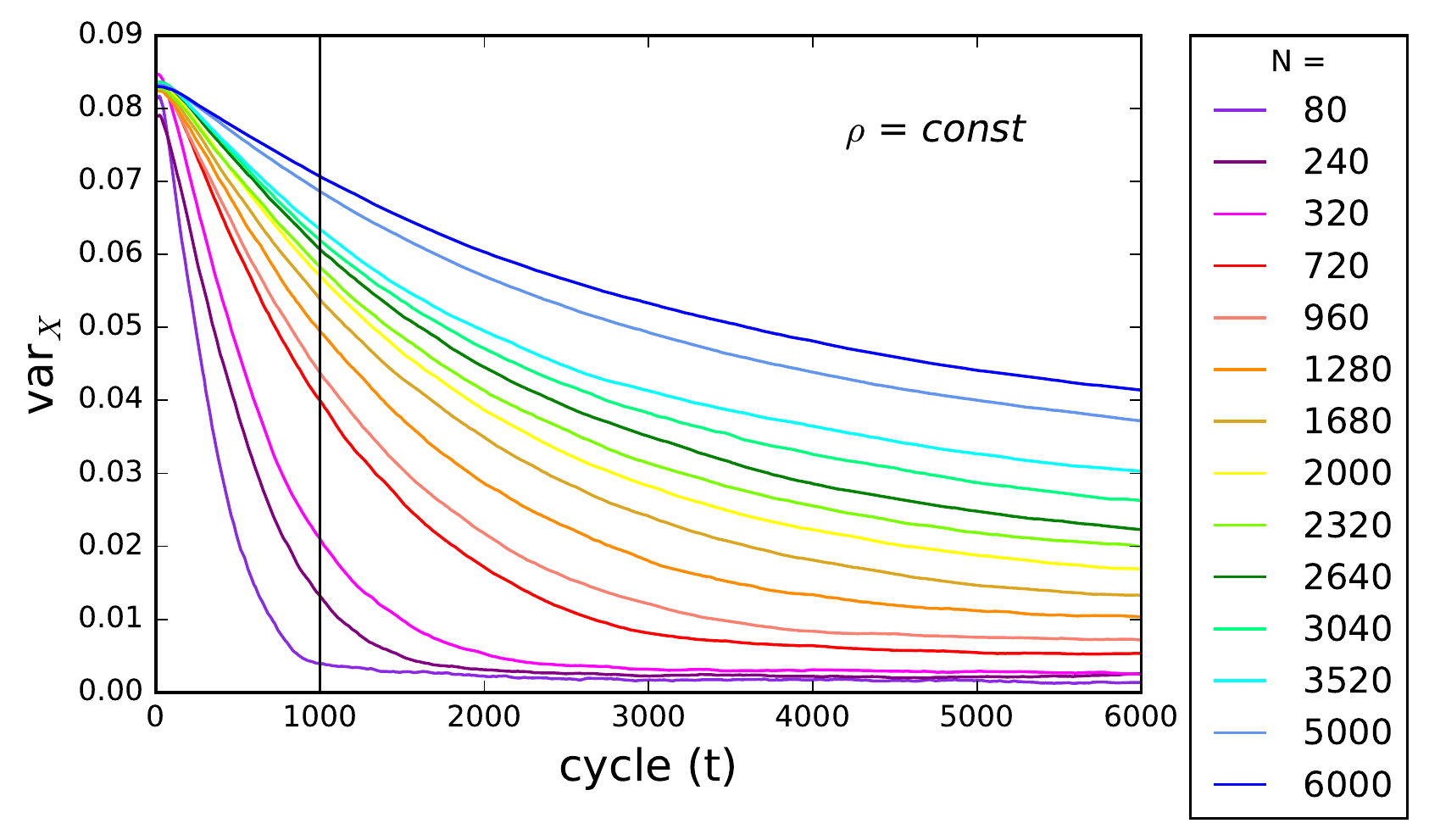}
  \caption{Several experiments at $\rho$ = const, thus varying $N$. Each colored curve depicts var$_{X}^{k}$ vs $t$, whose trend is exponential. The vertical line corresponds to $t = 1000$ so that, for each experiment $k$, $\mathrm{var}_{X}^{k}$(t=1000) is given by the intersection between such line and the $k$-th curve. Note that, for clarity reasons, $\sigma_{\mathrm{var}_{X}^{k}}$ is not shown.}
 \label{3}
 \end{figure}

After having obtained (var$_{X}^{k}(t) \pm \sigma_{var_{X}}^{k}(t)$) for each k-th experiment (that is, each choice of physical variables), they were compared by focusing on the very first stages of the aggregation process. Since var$_{X}$ vs $t$ follows an exponential rule: 
\begin{equation}
\mathrm{var}_{X}(t) = C + A\cdot e^{-B\cdot t}
\label{expo}
\end{equation}
it was informative to study how the transient behavior towards the aggregate differs quantitatively from experiment to experiment. Therefore, the interval of interest is that in which $| \frac{d \mathrm{var}_{X}}{d t}| > 1$, namely where the exponential trend is stronger. As stated previously, when \textit{in vitro} experiments are concerned, biologists begin to see a ``visual aggregate'' after approximately 6h \cite{fey}, corresponding to circa 400 cycles in our simulation. Hence, such non-informative lag was ignored; additionally, by doing so, it was possible to neglect such a transient whose characteristics are mainly due to the initial conditions. Thus we focused on the interval $\mathbb{I}= \{t: 400 < t < \tilde{t}\}$, where $\tilde{t}$ is such that $|\frac{d \mathrm{var}_{X}}{d t}\Bigr|_{\substack{t=\tilde{t}}}| \leq 1$. The value $t = 1000$ was then conventionally chosen as the representative for $\mathbb{I}$; consequently, what is going to be studied to estimate the quantitative differences between each $k$-th experiment is $\mathrm{var}_{X}^{k}$(t=1000). An example of the process of defining $\mathrm{var}_{X}^{k}$(t=1000) is seen in Fig. \ref{3}.\\

\section{Results}
To begin with, it was assessed the impact of the variables that define the model scale on the dynamics, namely $\rho$ (macroscopic) and $N$ (microscopic).
In all the experiments we chose both the cell food source and the amoebas to be uniformly distributed over the environment.

\subsection{$N=\mathrm{const}$}
\begin{figure}[h!]
  \centering
  \includegraphics[scale=0.45]{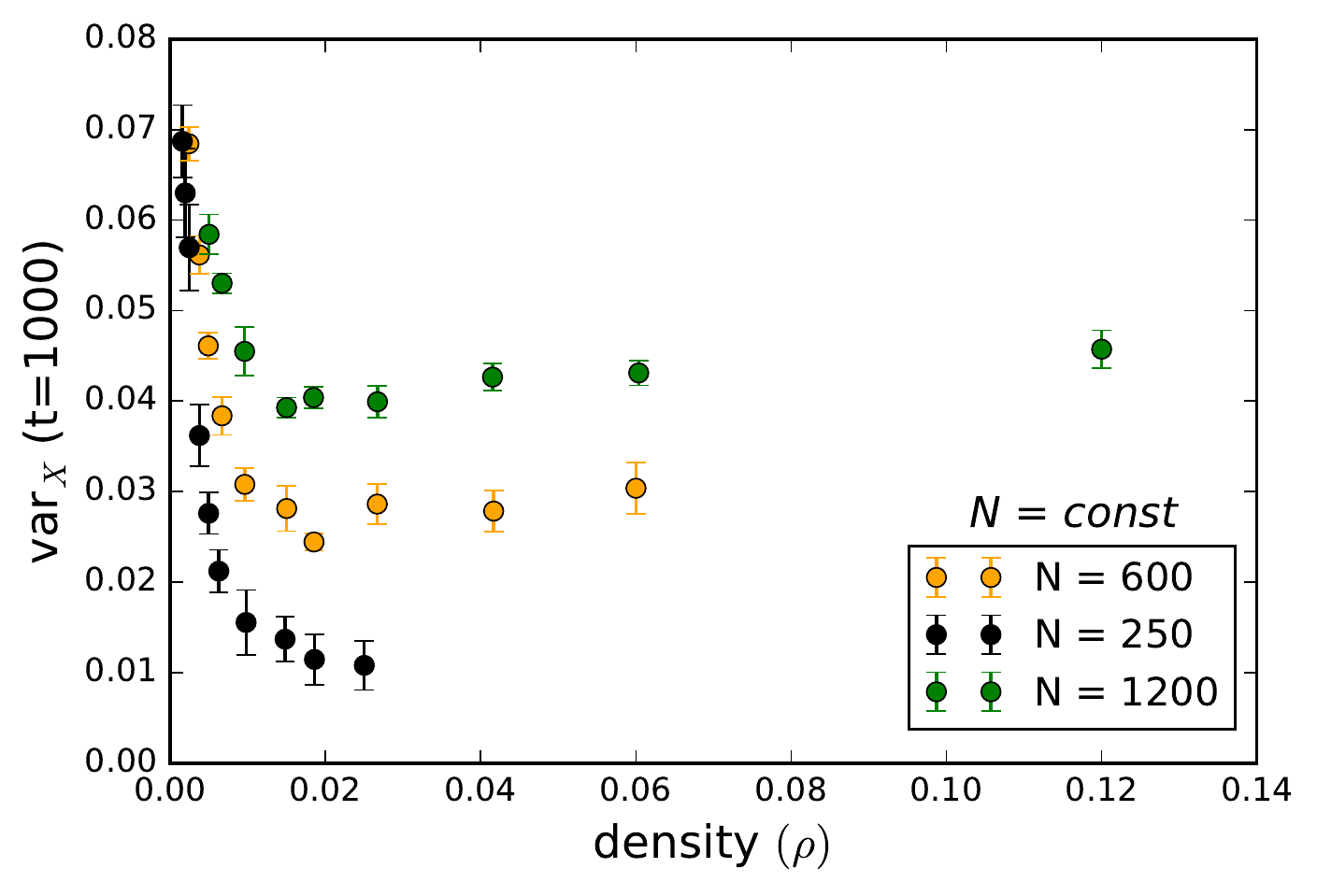}
  \caption{Three set of experiments, each of them characterized by a different $N$ value, are shown. Note that they show the same qualitative behavior. The statistical associated errors are evaluated as explained before.}
 \label{4}
 \end{figure}

The first one to be inquired is the macroscopic variable $\rho$, namely the density of amoebas. In order to do so, three different $N$ (respectively 250, 600, 1200) were selected and kept constant, thus defining three experimental sets; then the density for each k-th experiment was varied. As stated before, var$_{X}^{k}$(t=1000) was calculated as the quantitative estimator. Global results are shown in \ref{4}. \\

\noindent As it can be seen from the chart, we can heuristically say that there are two distinct dependences of var$_{X}$ on $\rho$.
For small values of $\rho$, var$_{X}$ decreases as the density increases, while for bigger values var$_{X}$ seems to "saturate" reaching a ``plateau''. Considering the chart, it is important to stress that: a) the critical density at which the functional dependence changes is not independent of $N$; b) the value of the "plateaus" of var$_{X}$ increases with $N$.  Not only that: what other studies called the ``typical density range''\cite{dallon}  seems to correspond to the ``saturation'' region only; we believe that our analysis, at least in the present model, have succeeded in highlighting the presence of a lower region in which the system shows a different behavior. \\
Finally, while in continuous models it was generally accepted that the most important variable was $\rho$ alone, it is evident that the dynamics of our MAS discrete system depends on the number of amoebas as well; hence, it is necessary to study the numerosity, looking for a scale-free range in which the dependence on it can be neglected.

\subsection{ $\rho = \mathrm{const}$}
\label{rhoconst}
\begin{figure}[h!]
  \centering
  \includegraphics[scale=0.4]{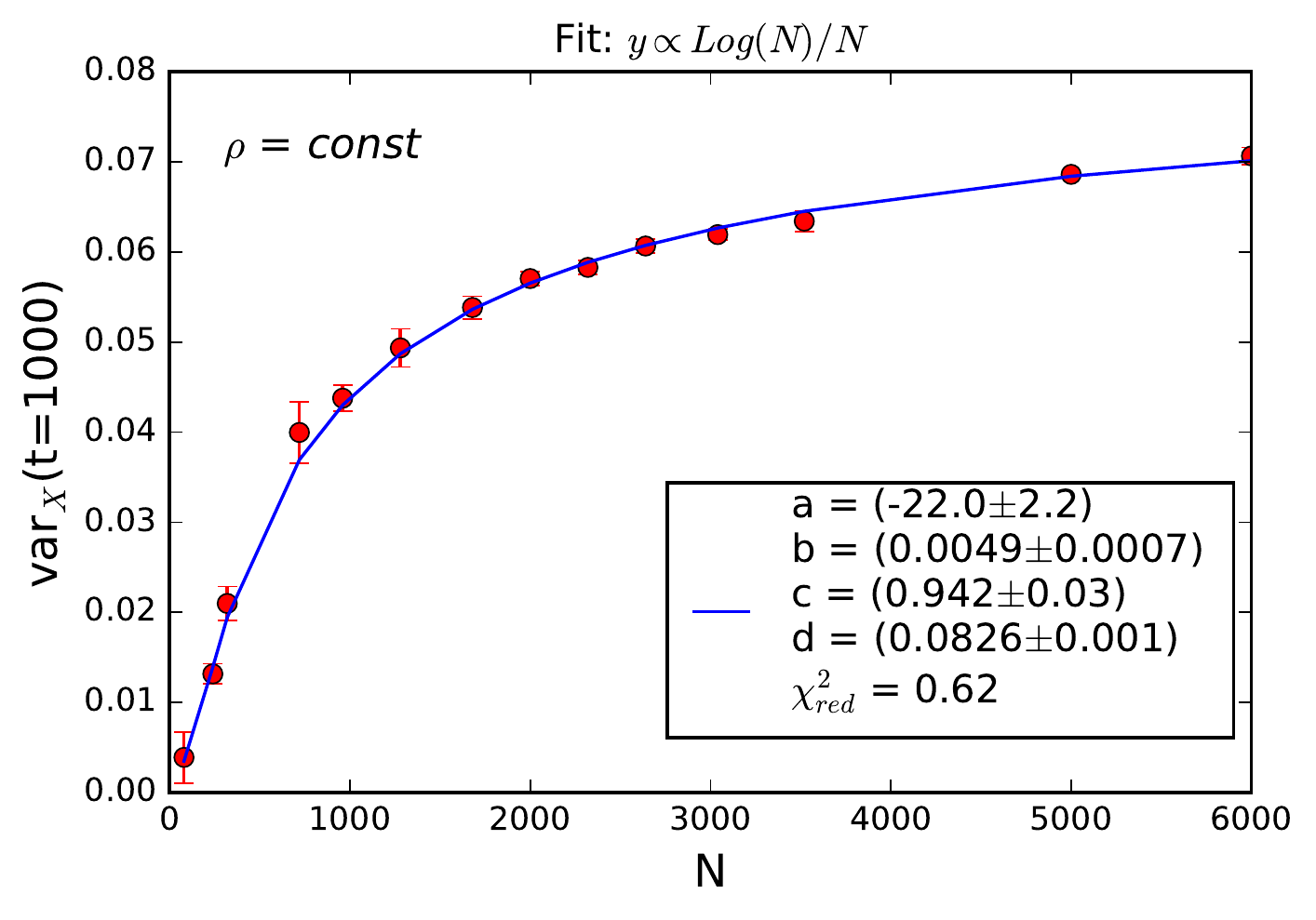}
   \caption{The quantitative estimator var$_{X}$(t=1000), as well as its associated statistical error, was calculated as described previously. As every experiment set showed the very same dependency, thus the same trend, only the one identified by the value $\rho = \frac{80}{100^{2}}$ is depicted.}
 \label{5}
 \end{figure}

In order to evaluate the impact of $N$, three different set of experiments were chosen, each of them with a specific constant density (respectively $\rho = \frac{80}{100^2}, \frac{120}{100^2}, \frac{200}{100^2}$). After that, $N$ was varied for each $k$-th experiment. The results are shown in Fig. \ref{5}.\\

\noindent As the $\chi^{2}$ guarantees, the fit follows 
\begin{equation}
\mathrm{var}_{X}(t=1000) = a \cdot \frac{\log(b \cdot N + c)}{N} + d
\label{slowlog}
\end{equation}
therefore resulting in a $\mathrm{var}_{X}(t=1000) \propto Log(N)/N$ trend. It was verified that all sets of experiments follow the very same rule, albeit with different values for the fit parameters. As a consequence, hereafter we focus on one particular experiments set, that is $\rho = \frac{80}{100^{2}}$.\\
The function (\ref{slowlog}) tends to an asymptote, so in principle it is possible to identify a scale-free range for the variable $N$; however, the value of $N$ for which var$_{X}(t=1000)$ differs by 5\% from the asymptotic value 25,000, which is a biological magnitude of numerosity. Despite its relevance for evolutionary studies, said scale was found not to be handleable with ease since it exceeds our own computational limits; future simulations with greater computation capabilities may exploit this result to set their variables in a purely scale-free region. Moreover, it may be interesting, for the purpose of scaling the model to lower values of $N$, to understand and quantify the error that comes from such scaling choice. Considering $\mathrm{var}_{X}(t=1000) = f(N)$ a proxy to identify an interval 
\[\mathbb{L}=\{\tilde{N}\, \mathrm{s.t.} \, 0< \frac{\partial f}{\partial N}\Bigr|_{\substack{\tilde{t}}} <1\ \, \mathrm{and} \, \tilde{N} \, \textrm{is computationally treatable}\}\]
we chose the value $\tilde{N}=2640$ to represent $\mathbb{L}$. Starting from that, it was possible to estimate the induced systematic error associated with MAS models that are placed in $\mathbb{L}$ instead of in a scale-free regime.\\

\subsection{Systematics}
\begin{figure}[h!]
  \centering
  \includegraphics[scale=0.45]{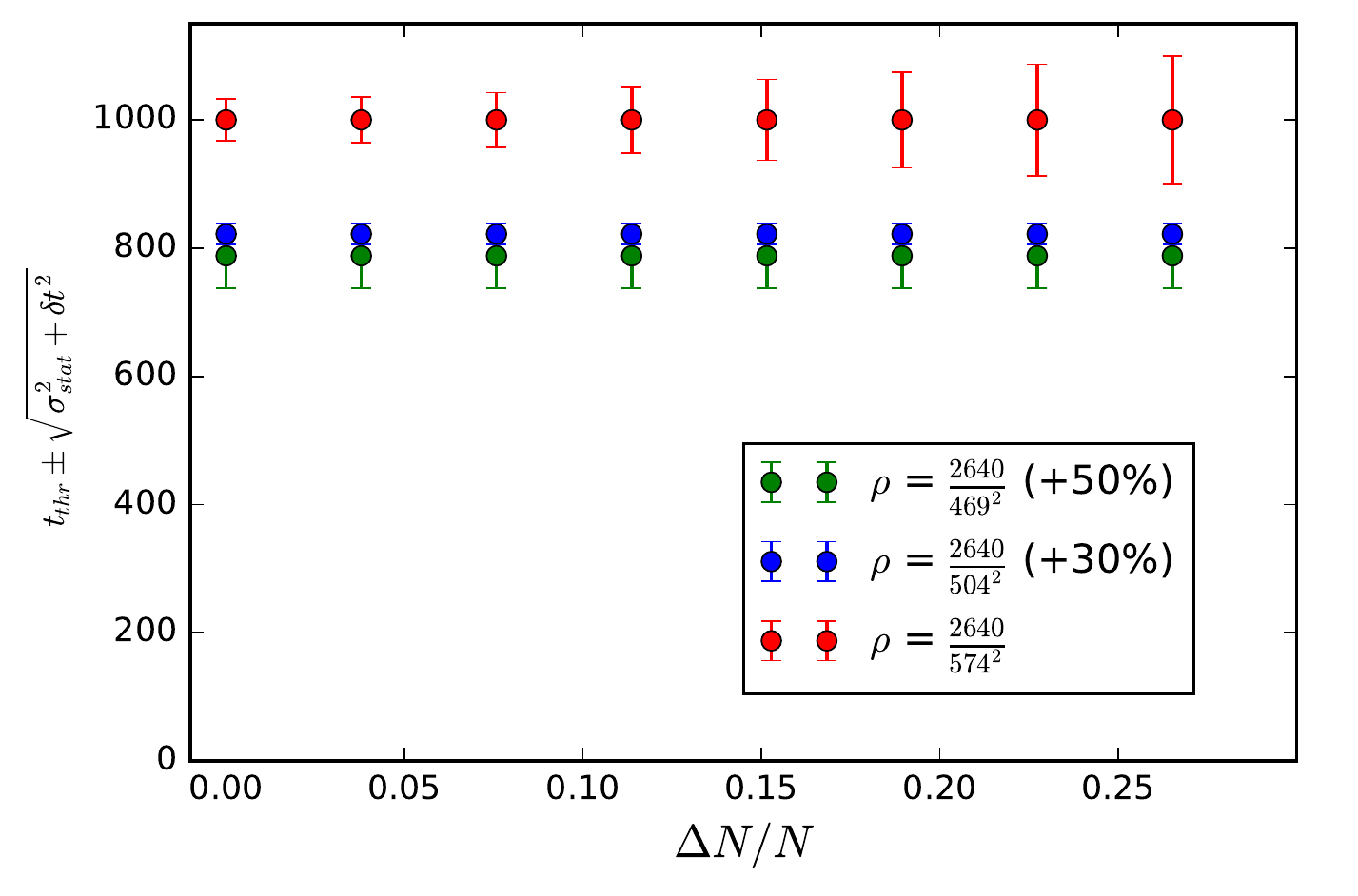}
   \caption{$t_{thr}$ vs  $\Delta N/N$. Three sets of experiments are shown and compared. The error bars are a superimposition of statistical and systematic error, so that it is possible to estimate how big the total error is in comparison with the sensitivity. Calling the critical density $\tilde{\rho}$, note that the two experiments with $\rho > \tilde{\rho}$ (namely the blue and green one) are already indistinguishable, as the are in a density-scale-free region, while the third one ($\rho < \tilde{\rho}$) can be statistically separated if and only if $\Delta N/N$ is ``low'', that is, if the associated systematics does not exceed a confidential value. Otherwise, a normal test Z cannot tell between the three sets.}
 \label{6}
 \end{figure}

The aforementioned systematic error was not determined on the estimator var$_{X}(t=1000)$ but directly on the order parameter $t$ instead. The reason is the following: in a real biological experiment, one cannot control the evolution of the amoeba dispersion, which necessitates an extensive control over every single cell, but can easily measure the time needed to reach an aggregation state. Knowing the dependences given by eq. \ref{expo} and \ref{slowlog}, it is straightforward to obtain, in $\mathbb{I}$:
\begin{equation}
t(N) = -\frac{1}{B}\cdot \log\{\frac{\frac{a\cdot Log[b \cdot N + c]}{N}+ d - C}{A}\}
\end{equation}
(recall that it was verified that $t=1000$ is a valid proxy for every $t \in \mathbb{I}$) . \\After that, once the desired aggregation state (defined by a value $\tilde{\mathrm{var}_{X}}$) is chosen, one can measure $t_{thr}$, namely the time that is needed to reach it (Fig. \ref{6}).\\
After that, the systematic error is given by $\delta t(N,\Delta N)$.
To evaluate it, we assumed a range of uncertainty on the number of amoebas $\Delta N$; then we hypothesized a linear dependence of $\delta t$ on $\Delta N$ as it was an error propagation. Such hypothesis was verified \textit{a posteriori} by evaluating the values of $\delta t_{exp}$ coming from interpolation of experimental data; to do so, we selected several values of $\hat{N}$ around $N=2640$ to define a $\Delta N$; for each of them we evaluated the corresponding var$_{X}^{\hat{N}}(t=1000)$ using equation (\ref{slowlog}); finally, we estimated $\hat{t}$ such that var$_{X}^{\hat{N}}(t=\hat{t})=\mathrm{var}_{X}^{2640}(t=1000)$ by using equation (\ref{expo}). Thus
\begin{equation}
\delta t(N,\Delta N) = \frac{\partial}{\partial N}(-\frac{1}{B}\cdot Log\{\frac{\frac{a\cdot Log[b \cdot N + c]}{N}+ d - C}{A}\}) \Delta N
\end{equation}

\noindent In order to contrast the systematics against the sensitivity of the proposed model (so that one may estimate how much the magnitude of $\delta t$ impacts on the statistical analysis), three more experiments were performed, each of them being assigned a different value of density. This way we consider the sensitivity as the capability of discriminating different regimes depending on the macroscopic variable $\rho$. With reference to the previous results, the three values were chosen around the critical density which leads to ``saturation'': $\rho = \{\frac{80}{100^{2}}, \frac{104}{100^{2}}, \frac{120}{100^{2}}\}$ (of course, as $N$ of interest is $N=2640$, they were scaled accordingly). Thanks to such decision, it was possible to compare the sensitivity in and out the scale-free region of $\rho$.
Results are shown in Fig. \ref{6}. \\

\noindent There are three main identified features that emerges from such analysis.\\ First, it was confirmed that, in a density-scale-free region, the experimental sets defined only by the choice of $\rho$ are not distinguishable - even without considering the systematics; as a consequence, we can identify an \textit{intrinsic} $\rho$-scale-free region that does not depend on measurement performances.\\ Second, the experimental mutual choice of $N$ and $\Delta N$ has a significant impact on $\delta t$, that is, if we work with a number of amoebas which is lower than the biological ones, we ought to pay particular attention to its uncertainty. It is also true the opposite: when setting the system to biological numbers, $\Delta N$ may be tuned less accurately, as it doesn't impact too much on the aggregation trend; this idea may lead to biological suggestions like ``how the evolution tuned the colonies to large numbers that are minimally dependent on stochastic fluctuation''.\\ Third, it is clear that, when modeling a computational discrete simulation, one must pay attention not only to the macroscopic variable $\rho$, but to the microscopic $N$ as well. If not, in the case that one notices indistinguishable results, he couldn't say for sure if they were intrinsically so or, on the contrary, a systematic error shaded them. Therefore, when setting a computational model, there is not such a thing like a ``$\rho$-scale-free region'' or a ``$N$-scale-free region'' alone, but there is a ''$\rho$-$N$-scale-free surface'' where the experimentalists should set their variables, knowing that a certain error is associated to sets that are outside such surface (but not too far, as we stressed during the whole analysis).

\section{System robustness}
\begin{figure}[h]
\centering

\begin{minipage}[c]{0.45\textwidth}
\includegraphics[scale=0.3]{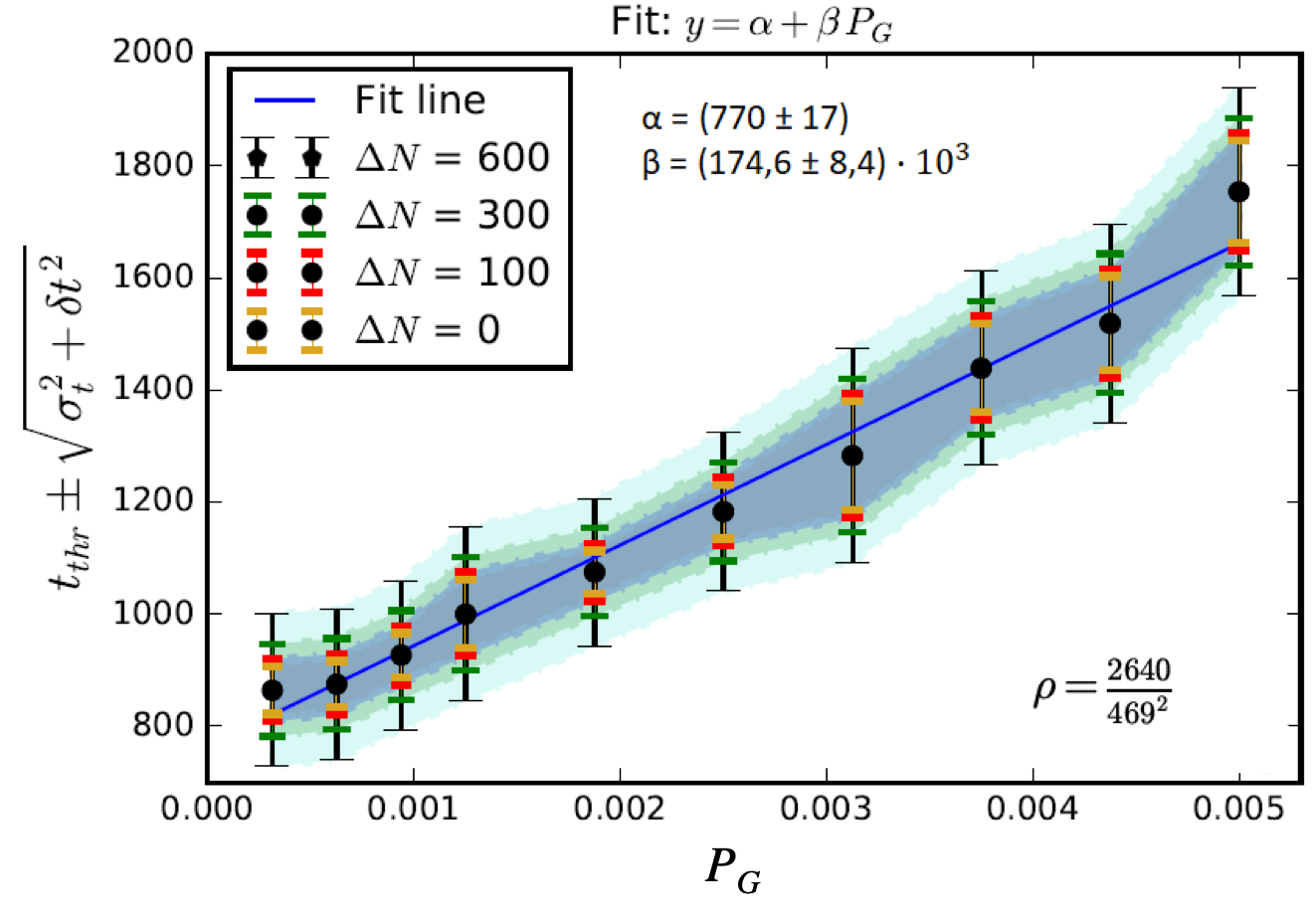}
\end{minipage}
\hspace{0.4cm}
\begin{minipage}[c]{0.45\textwidth}
\includegraphics[scale=0.45]{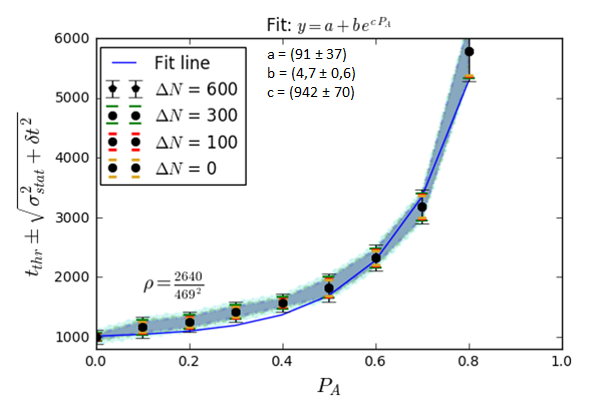}
\end{minipage} 
\hspace{0.4cm}
\caption{\small Experimental results for the dependence of the aggregation time on $P_{G}$ (left) and $P_A$ (right). Along with the mean values, four different error bars are shown: the statistical alone (with $\Delta N =0$), and three superimposed systematic, each of them depending on a different $\Delta N$. With regard to $P_A$, the linear fit is guaranteed by $\chi^2_{red} < 1$ for every associated error. As for $P_A$, despite the clear exponential trend, $\chi^2_{red} < 1$ only with a $\delta t(\Delta N = 600)$ associated error. Once again, we see that the associated systematics is not a formal, \textit{a priori} accident, but it is a necessary and informative quantity that must be taken into account when working within the considered parameter surface.}
\label{7}
\end{figure}

In biological systems, robustness is defined as the ability to maintain functionality in the presence of internal and external perturbations \cite{kitano2004biological,stelling2004robustness}; uncovering its mechanisms is a key issue in system biology \cite{kitano2007towards}. In the present context, it is of great importance to study the ``thermal-noise'' parameters, namely $P_G$ and $P_A$ (Sec. \ref{params}).\\
In consistency with the analysis conducted so far, we set $N=2640 \in \mathbb{L}$ and $\rho = \frac{2640}{469^2}$ (in the ``density-scale-free'' region). For the reasons stated before, it was estimated $(t_{thr} \pm \sigma_{stat})$ as well as the associated systematic error $\delta t$ for different values of $\Delta N$ (conventionally, a big a medium and a small one: $\Delta N$ = \{100, 300, 600\}). Results of $(t_{thr} \pm \sqrt{\sigma_{stat}^{2}+\delta t^{2}})$ vs $P_{G}$ and vs $P_A$ are shown in Fig. \ref{7}.

\noindent Regarding the role of $P_G$, big values of $\Delta P_{G}$ have to be chosen in order to magnify the linear trend (in fact, if $P_{G} = 0.00125$ is the reference value, the study interval is $\mathbb{P}=\{-75\% ; +300\%\}$). Therefore, large perturbations are necessary to significantly increase the aggregation time. On the other hand, $P_A$ leads to a possible critical slowing down of the system \cite{scheffer2009critical} towards the loss of attractor when values are too high. However, the aggregation rate is nonetheless consistent to biological values up to $P_A \simeq 0.5$, meaning that the system as a whole is robust even in adverse cases when individuals can fail half of the times to correctly process the chemical signal (inner noise). In other words, non-linearity of the problem balances different sources of noise, leading to controlled mean collective behaviors.

\section{Validating model predictions}
The model makes two important preliminary predictions about the colony evolution: (i) a finite set of individual behavioral rules, coupled with social interactions, are sufficient to elicit self-organization and decentralized gathering; (ii) thanks to mutual interactions, the system is robust against numerous sources of stochasticity. The first prediction has already been shown during the whole text, as it is the main verified hypothesis that led to the model design and implementation; however, we believe that, by expanding the class of candidate models, we may be able to produce an even better model. As for the second prediction, not only it has already been shown how the system collective behavior is robust against biological-like smearing of trajectories ($\delta_1$, $\delta_2$), but ``thermal-noise'' sources has been studied, too. As a result, we showed how the system is able to cope with internal and external perturbations, evolving in its original basin of attraction until said perturbations became greater than those found in nature.\\
To summarize, all the results (comparison with literature results after having set the same initial conditions, comparison with biological knowledge, assessment of a scale-free region for simulation repeatability and system robustness, validation of predictions) concur in the model validation, as per guidelines followed by other simulations \cite{bhall,khan}.

\section{Conclusions}
A \textit{Multi-agent system} approach could represent an innovative methodology to tackle complex biological systems such as the \textit{D. discoideum} aggregation process. In the present study we firstly validated our model. To begin with, it was identified a ``scale-free surface'' in which setting repeatable and reliable simulations; likewise previous research, which speaks of a ``normal density region'', we highlighted the presence of a $\rho$-scale-free-region towards which the system tends when values are biologically compatible. Not only that: by studying the number of individual cells, we demonstrated that, at least in a MAS approach, the macroscopic variable $\rho$ alone is not sufficient to fully set the model, but that a researcher needs to care about the microscopic variable $N$ as well. As a consequence, we highlighted a $\rho$-$N$-scale-free surface in which the experimentalist should place his variables during simulations. However, since such region is identified by values that are computationally hard to manage, we analyzed its neighborhood for $N$ slightly off-scaled ($\tilde{N}$, see Sec.\ref{rhoconst});  after that, we suggested how to consider the induced systematic error. This analysis states that it is important to handle an eventual uncertainty over $N$ with care, as it may lead to ambiguity during the comparison of different set of variables and/or hypothesis. On the contrary, if $\Delta N$ is treated adequately, the present analysis shows how to obtain quantitative results for a rescaled model to smaller, non-biological values of numerosity, easier managed by an \textit{in silico} simulation. Finally, the validated model was used to perform predictions about the system robustness against internal and external perturbations, finding interesting results about the system control.\\
The present model focuses on the first stages of \textit{D. discoideum} aggregation patterns; however, an interesting future perspective may be to extend the class of possible models by going beyond the present hypothesis. Other than that, this model can represent a suitable testbed to study dynamical and topological properties of the system. Moreover, we believe that the present model could have interesting practical applications in the field of system control and self-synchronization of robotic populations \cite{vlass,sugihara,dorigo,hart}.\\
To end with, we suggest that our project could represent a link between an abstract and theoretical mathematical modeling and a biological and more qualitative description because, on one hand, the replica of the physical space (``wet lab-alike simulation'') guarantees an high level of interpretability; on the other hand, the extensive control of the parameters and the possibility to realize quantitative measures and repeated tests allows going beyond the experimental limits of the \textit{in vivo} or \textit{in vitro} observations.\\

\section*{Acknowledgments}
The authors would like prof. P. Terna and dr. J. Markdahl for valuable feedbacks.

\nocite{*}

\bibliography{bibliography}
\bibliographystyle{unsrt}

\end{document}